# Green light GaN p-n junction luminescent particles enhance the superconducting properties of B(P)SCCO Smart Meta-Superconductors (SMSCs)

Qingyu Hai, Honggang Chen, Chao Sun, Duo Chen, Yao Qi, Miao Shi and Xiaopeng Zhao *

Smart Materials Laboratory, Department of Applied Physics, Northwestern Polytechnical University, Xi'an 710129, China;
* Correspondence: xpzhao@nwpu.edu.cn

**Abstract:** Superconducting materials exhibit unique physical properties and hold great scientific value and vast industrial application prospects. However, due to limitations such as critical temperature ($T_C$) and critical current density ($J_C$), the large-scale application of superconducting materials remains challenging. Chemical doping has been a commonly used method to enhance the superconductivity of B(P)SCCO. However, satisfactory enhancement results have been difficult to achieve. In this study, we introduced green light GaN p-n junction particles as inhomogeneous phases into B(P)SCCO polycrystalline particles to form a smart meta-superconductors (SMSCs) structure. Based on the electroluminescence properties of the p-n junction, the Cooper pairs were stimulated and strengthened to enhance the superconductivity of B(P)SCCO. Experimental results demonstrate that the introduction of inhomogeneous phases can indeed enhance the critical temperature $T_C$, critical current density $J_C$, and complete diamagnetism (Meissner effect) of B(P)SCCO superconductors. Moreover, When the particle size of raw material of B(P)SCCO is reduced from 30μm to 5μm, the grain size of the sintered samples also decreases, and the optimal doping concentration of the inhomogeneous phases increases from 0.15 wt.% to 0.2 wt.%, further improving the enhancement of superconductivity.

## 1. Introduction

Superconducting materials have attracted widespread attention in the scientific community due to their unique physical properties, such as zero electrical resistance [1] and complete diamagnetism [2]. However, the low current carrying density and the requirement for low-temperature environments during operation limit the practical applications of superconducting materials to some extent. Therefore, the discovery of new superconducting materials and the continuous improvement of critical temperature ($T_C$) and critical current density ($J_C$) have been the ongoing goals in the scientific community in order to achieve room-temperature superconductors. Since the 1980s, with the discovery of cuprate [4,4] and iron-based superconductors [5-7], the critical temperature ($T_C$) has surpassed the McMillan limit (40K) and reached the liquid nitrogen temperature range (70K). The development of superconducting materials has entered the high-temperature superconductivity phase, greatly inspiring researchers' confidence. So far, nearly a hundred types of superconducting materials have been discovered, but the progress in increasing the critical temperature ($T_C$) remains slow.

In recent years, researchers have also discovered some emerging superconducting materials. Such as high-pressure superconductors [8-11], photoinduced superconductors [12-14], 2D superconductors [15-17], interfacial superconductors [18-20] and nickel-based superconductors [21-23]. In 2019, a team from Stanford University in the United States discovered superconductivity in $Nd_{0.8}Sr_{0.2}NiO_2$ thin film material, with a critical temperature ($T_C$) of approximately 15K [21]. In 2021, Xie et al. found superconductivity at the interface of $KTaO_3$ and $LaAlO_3$ crystals, with a critical temperature ($T_C$) around 2.2K, and achieved the transition between superconducting and insulating phases under electric field control [23]. In July 2023, Wang Meng et al. from Sun Yat-sen University reported the emergence of superconductivity in $La_3Ni_2O_7$ single crystal samples under a pressure of 14GPa, with a critical temperature ($T_C$) of 80K [23]. These superconducting materials cannot be further applied at the moment, but they greatly help us to further explore superconducting materials and investigate the mechanisms behind the generation of superconductivity.

Although progress in increasing the critical temperature ($T_C$) of superconducting materials has been slow, we still haven't achieved low-cost superconductors. However, with the development of science and technology, superconducting materials are playing an increasingly important role in fields such as electric grids [24], energy [25], healthcare [26], controlled nuclear fusion [27], and quantum computing [28-30]. Based on the zero-resistance property of superconductors, high-temperature superconducting wires like BSCCO, YBCO and REBCO have been made into superconducting cables for power transmission [31-33]. Nb-based alloys, known for their high critical current density ($J_C$), high critical magnetic field ($H_C$), and low manufacturing cost, are used to produce superconducting magnets with low Joule and magnetic losses, serving as important components in nuclear magnetic resonance imaging devices, tokamak devices or superconducting rotating machines [34,35]. Iron-based superconductors, with their high $H_C$ characteristics, will also play an increasingly important role in high magnetic field applications [36,37].

BSCCO superconductor is a typical copper oxide superconductor, consisting of three coexisting superconducting phases: Bi-2201 phase ($T_C$=20K), Bi-2212 phase ($T_C$=87K), and Bi-2223 phase ($T_C$=110K), all of which have a cubic crystal structure [38-41]. This structure can be seen as a pseudo-perovskite layered structure, where the $CuO_2$ layers serves as the intrinsic superconducting layer and the BiO layers, SrO layers, and Ca atomic layers alternate as charge reservoir layers [42,43]. The charge reservoir layers generate charge carriers, which transfer to the superconducting layer at the interface, thus maintaining the purity of the superconducting layer to some extent and reducing impurity scattering, facilitating macroscopic superconductivity. Furthermore, this structure can be viewed as a natural Josephson junction, and it has been proposed by researchers that BSCCO superconductors can be used to fabricate terahertz wave emitters based on the ac Josephson effect [44-46]. BSCCO superconductors have the characteristics of not containing rare-earth elements, high critical temperature ($T_C$), insensitivity to oxygen, and simple manufacturing process, making them the most widely used and promising superconducting materials in industrial applications. Enhancing the superconducting properties of BSCCO superconductors can not only further enhance the application value of BSCCO but also contribute to our exploration and understanding of superconducting mechanisms. So far, the most common method to enhance the superconducting properties of BSCCO is chemical doping, such as doping elements like Pb [47], Al [48], Cs [49], Ce [50] into BSCCO superconductors to improve their performance through element substitution. Studies have shown that in the case of Pb doping, where $Pb^{2+}$ partially replaces $Bi^{3+}$, the structure of the Bi-2223 phase becomes more stable and the content of the Bi-2223 phase is increased, but the critical temperature ($T_C$) cannot be improved. Similarly, doping with Al, Cs, Ce, and other elements cannot increase $T_C$, but it can effectively enhance the critical current density ($J_C$) at proper doping concentrations. Alternatively, doping compounds like $Al_2O_3$ [51], $SnO_2$ [52], $ZrO_2$ [53], MgO [54] into BSCCO superconductors cannot increase $T_C$, and in some cases, it may even decrease $T_C$, but at appropriate doping concentrations, $J_C$ can be effectively enhanced. Thus, it can be seen that the results of chemical doping methods are not ideal and cannot effectively increase $T_C$ and $J_C$.

Metamaterial is a kind of composite material with an artificial structure, allowing for the realization of unique or even non-existent "anomalous" material properties in nature [55-57]. Based on metamaterial methods, our research team has proposed a new type of smart superconducting material called Smart Meta-Superconductors (SMSCs). Superconducting polycrystalline particles are used as the matrix, and electroluminescent particles are introduced as the inhomogeneous phase into the superconducting polycrystalline particles, forming the structure of SMSCs. In previous studies, we introduced $Y_2O_3:Eu^{3+}$ and $Y_2O_3:Eu^{3+}$+Ag electroluminescent materials as the inhomogeneous phase into $MgB_2$ [58-63] and B(P)SCCO [64-66] superconductors, forming the smart meta-superconductors. We believe that this structure can treat superconducting particles as microelectrodes. When a current is applied, the electroluminescence of the inhomogeneous phase is excited by the electric field between the microelectrodes, stimulating and enhancing Cooper pairs, thereby improving the superconducting properties. Research has shown that this type of SMSCs material, which can be modulated and improved in superconducting properties by external field stimulation, has new characteristics that traditional chemical doping methods cannot achieve. When measuring the $T_C$ of SMSCs using the four-probe method, the external electric field stimulates the electroluminescence of the inhomogeneous phase (EL), thereby enhancing the Cooper pairs and increasing the critical temperature $T_C$. However, $Y_2O_3:Eu^{3+}$ and $Y_2O_3:Eu^{3+}$+Ag electroluminescent materials require strong electric field stimulation, and the low and unstable luminous intensity easily decays, which cannot effectively enhance the critical temperature $T_C$. In recent years, our team has also introduced red AlGaInP p-n junction particles as the inhomogeneous phase into conventional $MgB_2$ superconductors, forming smart meta-superconductors [67]. The results show that compared to $Y_2O_3:Eu^{3+}$ and $Y_2O_3:Eu^{3+}$+Ag electroluminescent materials, p-n junction particles can more effectively enhance the critical temperature $T_C$, critical current density $J_C$, and complete diamagnetism (Meissner effect) of $MgB_2$ superconductors due to their structural stability, high luminous intensity, and the ability to be activated at low voltage.

The successful incorporation of red AlGaInP p-n junctions as inhomogeneous phases into conventional superconductor $MgB_2$ to form smart meta-superconductors raises the question of whether it is equally effective for high-temperature superconductors. In this paper, we investigate the doping of green GaN p-n junction particles as inhomogeneous phases into high-temperature superconductor B(P)SCCO, forming smart material superconductors. The study demonstrates that the doping of p-n junction nanostructures ensures the stability of material performance and effectively enhances the critical temperature $T_C$, critical current density $J_C$, and complete diamagnetism (Meissner effect) of B(P)SCCO superconductors. Furthermore, as the particle size of the B(P)SCCO polycrystalline particles decreases with the reduction of raw material particle size, the doping content of the p-n junction inhomogeneous phase increases, further amplifying the enhancement effect on the superconducting properties.

## 2. Model

Figure 1 illustrates the concept of using metamaterials to construct a smart meta-superconductor model, based on polycrystalline B(P)SCCO as the matrix. The gray rectangles in the image represent polycrystalline B(P)SCCO particles. The green particles are GaN p-n junction luminescent particles that emit green light waves with a central wavelength of 550nm, and they are dispersed as inhomogeneous phases around the polycrystalline B(P)SCCO particles. When measuring the sample's performance, under the influence of an external electric field, the superconductor polycrystalline particles serve as microelectrodes, effectively exciting electroluminescence in the inhomogeneous phase, thereby stimulating Cooper pairs and enhancing superconducting properties. In addition, as the size of B(P)SCCO polycrystalline particles decreases, more inhomogeneous phases can be accommodated within the same spatial volume, further increasing the enhancement effect of superconductivity.

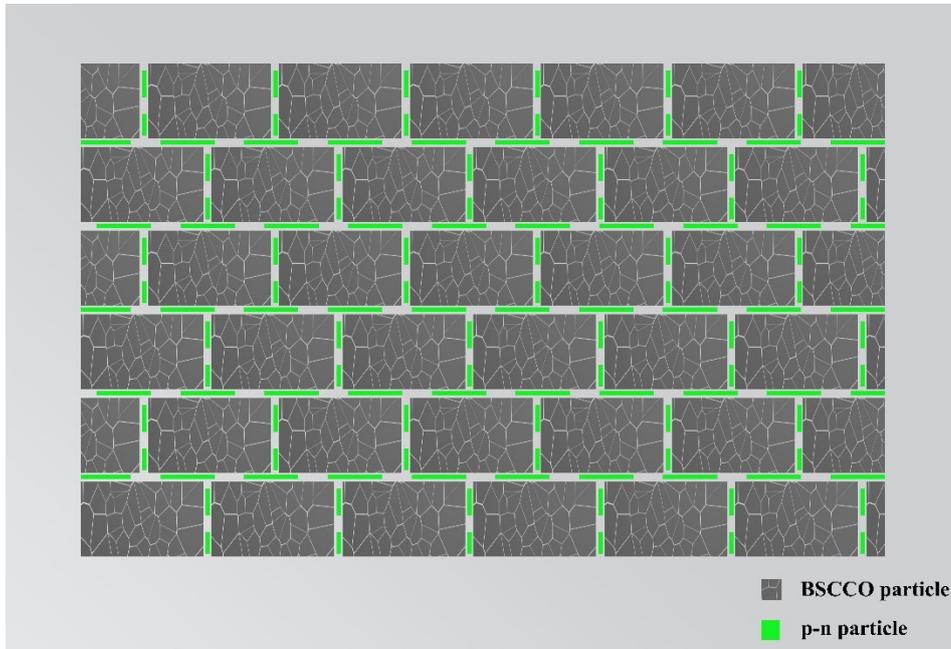

**Figure 1.** Diagram of B(P)SCCO SMSCs Model.

## 3. Materials and Methods

3.1. Preparation of p-n junction luminescent particles

The raw material used is commercial green light LED epitaxial wafer (Xiamen Qianzhao Optoelectronics Co., Ltd.), with GaN as the main component and an emission wavelength of 458nm. It was grown on a sapphire substrate using metal-organic chemical vapor deposition method, with the activation temperature of the p-type layer exceeding 1300°C. The p-n junction luminescent structure was detached from the $Al_2O_3$ substrate and mechanically ground to obtain p-n junction luminescent particles measuring 2μm × 2μm × 1.7μm as shown in the SEM diagram in Figure 2(a). The p-n junction particles ground from the substrate has a three-layer nanostructure of n-type GaN semiconductor, multi-quantum well and p-type GaN semiconductor. GaN is a stable compound with a melting point of 1700°C, so the p-n junction particles can withstand long-term annealing at 840°C. The GaN p-n junction particles can emit light when a voltage of <10V and a current of <10mA are applied, so during sample testing, the testing current is used to illuminate the p-n junction particles. Figure 2(b) shows the brightness graph of the p-n junction particles illuminated at a voltage of <10V and a current of <10mA. The green curve represents the brightness of the p-n junction particles, while the red curve represents the brightness of $Y_2O_3:Eu^{3+}$+Ag rare-earth oxide electroluminescent particles prepared by our team [68,69]. It can be seen from the graph that the brightness of the p-n junction luminescent particles is much higher than that of the $Y_2O_3:Eu^{3+}$+Ag rare-earth oxide electro-luminescent particles. Moreover, the p-n junction luminescent particles can maintain continuous luminescence for more than 1000 hours without brightness decay, indicating superior luminescent performance compared to the $Y_2O_3:Eu^{3+}$+Ag rare-earth oxide electroluminescent particles.

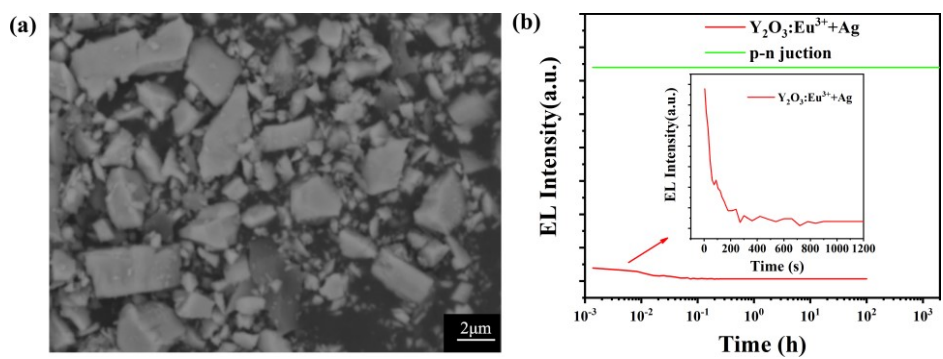

**Figure 2.** (a)SEM diagram of p-n junction particles, ground into 2 μm × 2 μm × 1.7 μm particles, (b) Luminescence intensity and lifetime test curves of p-n junction particles and rare earth oxide particles in red light wavelengths.

3.2. Preparation of B(P)SCCO precursor powders

A certain amount of $Bi_2O_3$ (Alfa Aesar, 99.99%), PbO (Alfa Aesar, 99.99%), $SrCO_3$ (Alfa Aesar, 99.99%), $CaCO_3$ (Alfa Aesar, 99.99%), and CuO (Alfa Aesar, 99.99%) powders were weighed according to the molar ratio of 1.6:0.4:2:2:3. The mixed raw powders were placed in a 50 ml ball milling jar with anhydrous ethanol as the grinding aid, and the ball milling machine was set at a speed of 500 rpm for 20 or 80 hours. The raw materials milled for 20 hours were dried and sieved through a 500mesh stainless steel sieve to obtain the raw material S1 with a particle size of about 30μm. The raw materials milled for 80 hours were filtered and dried using a G4 funnel, and then obtained the raw material S2 with a particle size of about 5μm. The raw materials S1 and S2 were separately placed in a tube furnace and annealed at 840°C in air for 50 hours + 50 hours. After the intermediate grinding, two different-sized black B(P)SCCO precursor powders were obtained.

3.3 Preparation of BSCCO Superconductor and Inhomogeneous Phase Sample

A certain amount of B(P)SCCO precursor powder was weighed and placed in a mold. The powder was pressed under a pressure of 12 MPa for 10 minutes to form a circular disk with a diameter of 12 mm and a thickness of 1 mm. The disk was then placed in a ceramic boat. The ceramic boat was transferred to a tube furnace and annealed at a temperature of 840°C for 30 hours. After cooling to room temperature, the disk was removed from the ceramic boat and thoroughly ground into powder. The ground B(P)SCCO powder and the corresponding inhomogeneous phase p-n junction particles with different mass fractions were placed in separate beakers and mixed with anhydrous ethanol to form a solution. The two solutions were then transferred to an ultrasonic cleaner and sonicated for 20 minutes. After sonication, the solutions were placed on a magnetic stirrer. During the stirring process, the inhomogeneous phase solution was added dropwise to the B(P)SCCO solution using a pipette. After completion, the mixture was stirred for 10 minutes and sonicated for 20 minutes. The mixed solution was then transferred to a petri dish and placed in a vacuum drying oven. The sample was dried at 65°C under vacuum for 5 hours to obtain a black powder. The dried B(P)SCCO black powder was placed in a mold and pressed under a pressure of 12 MPa for 10 minutes to form a circular disk with a diameter of 12 mm and a thickness of 1 mm. The disk was then placed in a ceramic boat. The ceramic boat was transferred to a tube furnace and annealed at a temperature of 840°C in an air atmosphere for 120 hours. After cooling to room temperature, the corresponding sample was obtained.

In the experiment, we obtained raw material S1 with a particle size of 30μm by ball milling for 20 hours and raw material S2 with a particle size of 5μm by ball milling for 80 hours. Samples made using raw material S1 are referred to as the A series, while samples made using raw material S2 are referred to as the B series. The sample numbers and doping concentrations of the A series and B series samples are shown in Table 1 and Table 2.

**Table 1.** Sample numbers and doping concentrations of the A series samples prepared using raw material S1 (30μm).

| Sample | A1 | A2 | A3 | A4 |
|---|---|---|---|---|
| Inhomogeneous phase p-n junction concentration（wt.%） | 0 | 0.1 | 0.15 | 0.2 |

**Table 2.** Sample numbers and doping concentrations of the B series samples prepared using raw material S2 (5μm).

| Sample | B1 | B2 | B3 | B4 | B5 |
|---|---|---|---|---|---|
| Inhomogeneous phase p-n junction concentration（wt.%） | 0 | 0.1 | 0.15 | 0.2 | 0.25 |

3.4. Measurement of critical temperature ($T_c$)

The resistance-temperature dependence (R-T) curves of the prepared samples at low temperatures were measured using the four-probe method to determine the zero-resistance transition temperature ($T_{C,0}$) and the onset transition temperature ($T_{C,\,on}$) of the samples. The entire testing process was conducted under vacuum conditions, with a closed-cycle cryostat provided by Advanced Research Systems to create the low-temperature environment (minimum temperature of 10K). The temperature during the testing process was controlled by a low-temperature controller from Lake Shore. The high-temperature superconducting material characteristic testing system produced by Shanghai Qianfeng Electronic Instrument Co., Ltd. provided the testing current (0.1-100 mA), and the voltage measurement was performed using a Keithley digital nanovoltmeter.

3.5. Measurement of Critical Current Density and Meissner Effect

Measurement of current-voltage (I-V) characteristics under zero magnetic field conditions was also performed using the four-probe method. The sample was connected to the leads using indium wire, and a certain value of current was applied to the sample through the outermost two leads, while the voltage was measured using a Keithley digital nanovoltmeter connected to the middle two leads. When the current flowing through the sample, denoted as I, exceeded a certain threshold, the superconducting state of the sample was disrupted and transitioned to the normal state. This critical current is known as the critical transport current of the superconductor. Typically, the critical transport current density ($J_C$) in a superconducting system is determined by measuring the I-V characteristics at different temperatures (below the onset transition temperature, $T_{C,\,on}$), with a voltage criterion of 1μV/cm. During the testing process, the shape and size of all samples, as well as the distances between the current and voltage leads, were kept constant. Subsequently, DC magnetization testing was performed on the prepared samples. The samples were slowly cooled in a 2.5mT magnetic field parallel to the plane, and data were collected during the subsequent warming process. All samples exhibited complete diamagnetism.

## 4. Results

4.1. Scanning electron microscope(SEM)

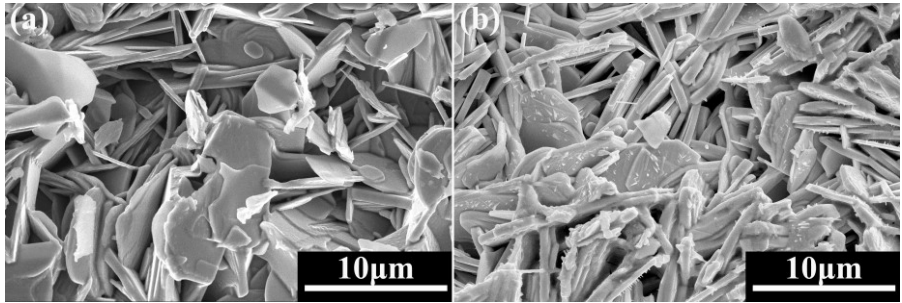

**Figure 3.** SEM diagram of the pure sample A1(a) and B1(b) prepared from the raw materials S1, S2 after sintering.

Figure 3 shows the SEM images of the pure B(P)SCCO samples A1 (Figure 3(a)) and B1 (Figure 3(b)) prepared from raw materials S1 and S2, respectively. From the images, it can be observed that all samples exhibit a randomly distributed layered structure. The layered structure of sample B1 is smaller in size and contains more gaps compared to sample A1. This indicates that as the size of the raw materials decreases, the particle size of the prepared samples decreases while the gaps between particles increase. Therefore, under the same volume, more p-n junction luminescent inhomogeneous phases can be accommodated. Figure 4(a) shows the SEM image of the B(P)SCCO sample (A3) doped with 0.15 wt.% p-n junction luminescent inhomogeneous phases. It can be observed that the introduction of p-n junction luminescent inhomogeneous phases does not seem to have a significant impact on the size of the sample's layered structure or the connectivity between particles. Figures 4(b-f) show the EDS mapping for elements Bi, Sr, Ca, Cu, and Ga listed in the top right corner of each figure. It indicates that the GaN p-n junction luminescent particles are randomly distributed around the B(P)SCCO layered structure.

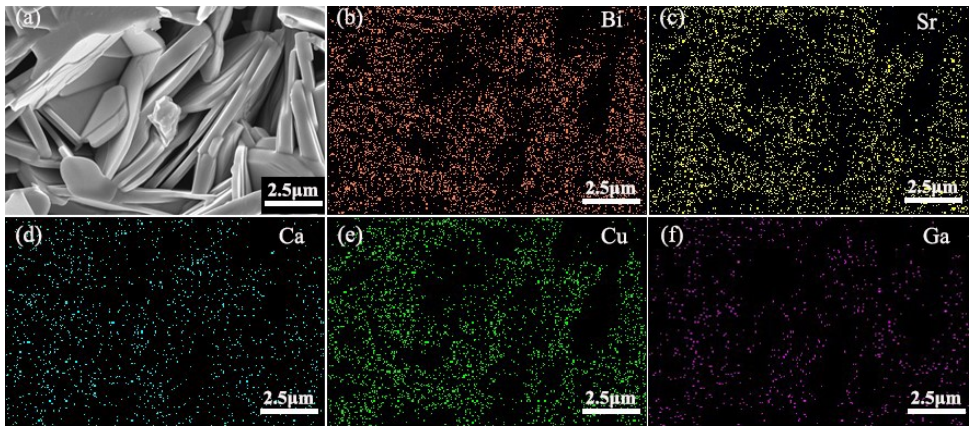

**Figure 4.** (a) SEM diagram of the sample(A3) after sintering, (b-f) EDS mapping of Bi, Sr, Ca, Cu and Ga.

4.2. Critical temperature $T_C$

Table 3 and Figure 5(a) show the critical temperature ($T_C$) and normalized R-T curves of the A-series samples prepared from raw material S1 with a particle size of 30μm, as measured by the four-probe method. The black curve in the figure represents the pure sample A1 of B(P)SCCO, with a transition temperature of 103-114K. The remaining three curves represent the inhomogeneous phase-doped samples A2, A3, and A4 with x wt.% (x=0.1, 0.15, 0.2) content, with transition temperatures of 103.5-114K, 105-114K, and 104-114K, respectively. Figure 5(b) shows the enhancement of the zero-resistance temperature (TC,0) in the A-series samples, denoted as $\Delta T_{C,0}$. It can be observed from the figure that the inhomogeneous phase doping effectively enhances the zero-resistance transition temperature TC,0 of the B(P)SCCO superconductor, and the enhancement $\Delta T_{C,0}$ shows a normal distribution relationship with the content of inhomogeneous phase doping. The optimal doping content is 0.15 wt.%, and the zero-resistance transition temperature $T_{C,0}$ is increased by 2K.

**Table 3.** Critical temperature ($T_C$) of A series samples

| sample | concentration (wt.%) | $T_{C,0}$/K | $T_{C,on}$/K | $\Delta T_{C,0}$/K | $\Delta T_{C,on}$/K | Transition Width/K |
|---|---|---|---|---|---|---|
| A1 | 0 | 103 | 114 | 0 | 0 | 11 |
| A2 | 0.1 | 103.5 | 114 | 0.5 | 0 | 10.5 |
| A3 | 0.15 | 105 | 114 | 2 | 0 | 9 |
| A4 | 0.2 | 104 | 114 | 1 | 0 | 10 |

**Table 4.** Critical temperature ($T_C$) of B series samples

| sample | concentration (wt.%) | $T_{C,0}$/K | $T_{C,on}$/K | $\Delta T_{C,0}$/K | $\Delta T_{C,on}$/K | Transition Width/K |
|---|---|---|---|---|---|---|
| B1 | 0 | 102 | 109 | 0 | 0 | 7 |
| B2 | 0.1 | 102.5 | 112 | 0.5 | 3 | 9.5 |
| B3 | 0.15 | 103 | 112 | 1 | 3 | 9 |
| B4 | 0.2 | 104 | 112 | 2 | 3 | 8 |
| B5 | 0.25 | 103.5 | 112 | 1.5 | 3 | 8.5 |

Table 4 and Figure 5(c) show the critical transition temperature ($T_C$) and normalized R-T curves of the B-series samples prepared from raw material S2 with a particle size of 5μm, as measured by the four-probe method. The gray curve in the figure represents the pure sample B1 of B(P)SCCO, with a transition temperature of 102-109K. The remaining four curves represent the inhomogeneous phase-doped samples B2, B3, B4, and B5 with x wt.% (x=0.1, 0.15, 0.2) content, with transition temperatures of 102.5-112K, 103-112K, 104-112K, and 103.5-112K, respectively. Figure 5(d) shows the enhancement of the zero-resistance temperature ($T_{C,0}$) in the B-series samples, denoted as $\Delta T_{C,0}$. It can be observed from the figure that the inhomogeneous phase doping effectively enhances the critical temperature $T_C$ of the B(P)SCCO superconductor, with an increase of 3K in the onset transition temperature $T_{C,on}$ and an enhancement $\Delta T_{C,0}$ in the zero-resistance transition temperature $T_{C,0}$. The enhancement $\Delta T_{C,0}$ also show a normal distribution relationship with the content of inhomogeneous phase doping. The optimal doping content is 0.2 wt.%, and the zero-resistance transition temperature $T_{C,0}$ is increased by 2K.

From Table 3, Table 4, and Figure 5, it can be seen that the critical temperature ($T_C$) of the A-series samples is higher than the $T_C$ of the B-series samples. This is because the grain size of the B(P)SCCO samples prepared from raw materials with different particle sizes is different. The grain size of the B-series samples is smaller than that of the A-series samples, and there are more grain boundaries between the polycrystalline particles in the B-series samples than in the A-series samples, resulting in poorer connectivity in the B-series samples and thus a decrease in the critical temperature $T_C$. However, the reduction in grain size and the increase in grain boundaries allow the B-series samples to accommodate more inhomogeneous phases. Therefore, we can see from the figure that the optimal doping concentration for the A-series samples is 0.15 wt.%, while the optimal doping concentration for the B-series samples is 0.2 wt.%, and the enhancement of the critical temperature $T_C$ in the B-series samples is better than that in the A-series samples.

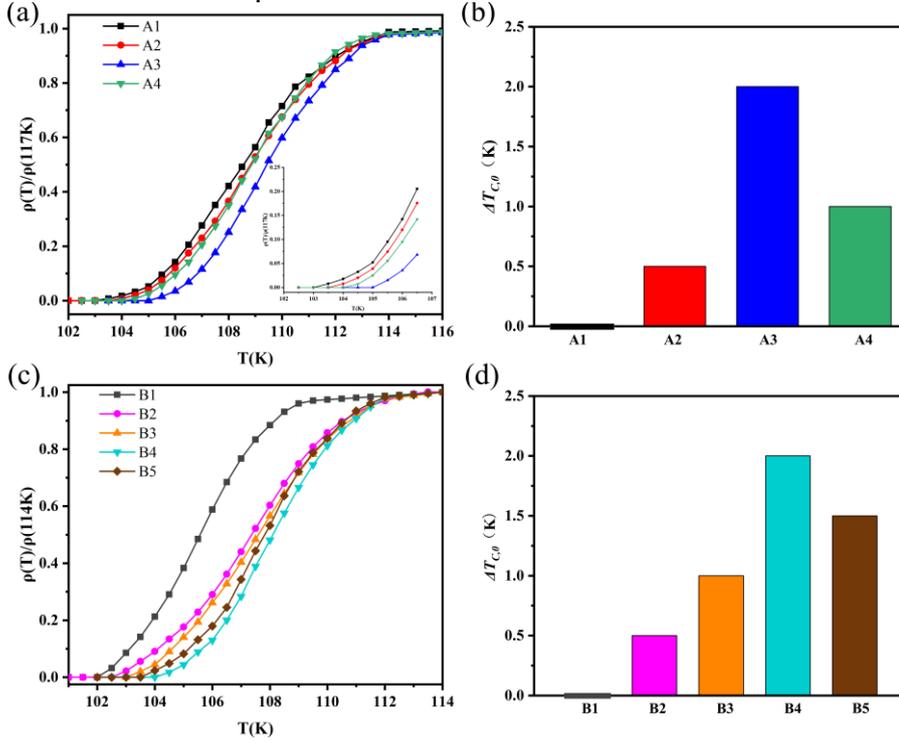

**Figure 5.** (a) Normalized R-T curve of A series samples; (b) Enhancement of zero-resistance transition temperature $T_{C,0}$ of A series samples, $\Delta T_{C,0}$; (c) Normalized R-T curve of B series samples; (d) Enhancement of zero-resistance transition temperature $T_{C,0}$ of B series samples, $\Delta T_{C,0}$.

*4.3. Critical current density $J_C$*

Figures 6(a) and (b) show the relationship between the critical current density ($J_C$) and temperature (T) for the pure B(P)SCCO sample A1 and the inhomogeneous phase-doped B(P)SCCO samples A2, A3, and A4 with x wt.% (x=0.1, 0.15, 0.2) content, prepared from raw material S1 with a particle size of 30μm. Figures 6(c) and (d) show the relationship between $J_C$ and T for the pure B(P)SCCO sample B1 and the inhomogeneous phase-doped samples B2, B3, and B4 with x wt.% (x=0.1, 0.15, 0.2) content, prepared from raw material S2 with a particle size of 5μm. These curves were obtained from I-V measurements. The critical current density $J_C$ of the pure B(P)SCCO samples at a temperature of 90K is 105 A/cm² and 56 A/cm², respectively, which is consistent with the literature. From the graphs, it can be seen that the $J_C$ decreases with increasing temperature, with a faster decrease in the low-temperature range and a slower decrease at higher temperatures. The $J_C$ decreases to a minimum at the onset transition temperature $T_{C,on}$, consistent with the relationship between the critical current $J_C$ and temperature T reported in the literature [70,71]. The increase in grain boundaries in the B-series samples prepared from smaller raw material sizes compared to the A-series samples leads to poorer connectivity, resulting in a lower critical current density $J_C$ for the B-series samples at the same temperature as the A-series samples. In addition, from the graphs, it can be seen that at the same temperature, the critical current density $J_C$ of the inhomogeneous phase-doped samples is higher than that of the pure B(P)SCCO samples. The optimal doping content for the inhomogeneous phase in the A-series samples is 0.15 wt.%, and at a temperature of 90K, the critical current density is increased by 25% compared to the pure B(P)SCCO sample. The optimal doping content for the inhomogeneous phase in the B-series samples is 0.2 wt.%, and at a temperature of 90K, the critical current density $J_C$ is increased by 35% compared to the pure B(P)SCCO sample. This indicates that inhomogeneous phase doping can effectively enhance the critical current density $J_C$ of B(P)SCCO superconductors, and the smaller the grain size, the higher the content of inhomogeneous phase that can be accommodated, resulting in better enhancement.

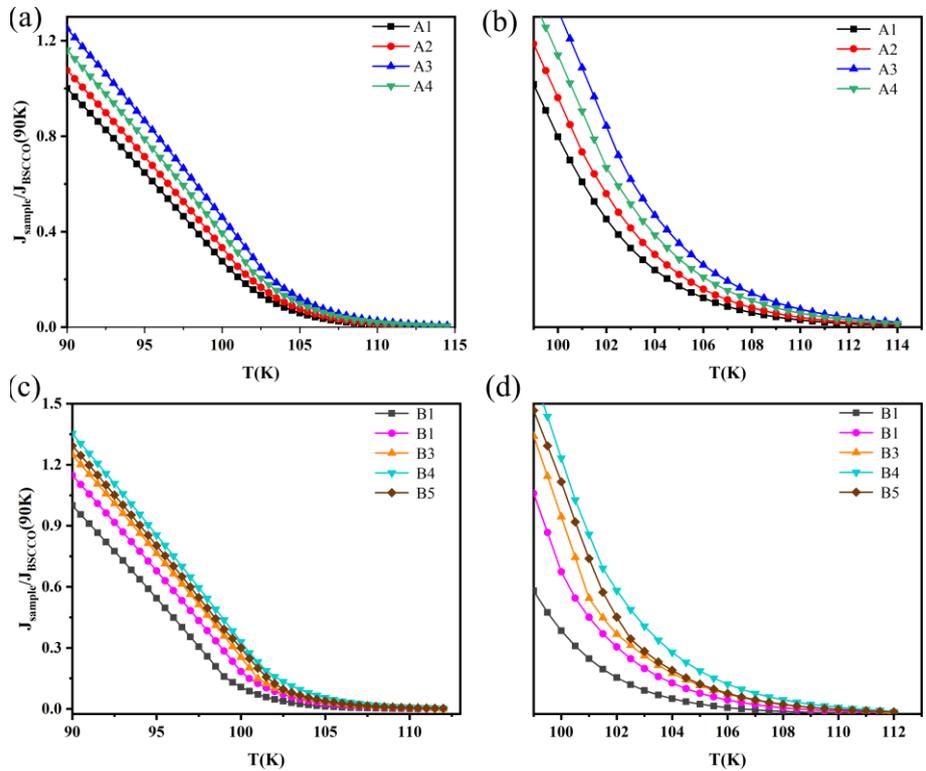

**Figure 6.** The relationship between critical current density ($J_C$) and temperature for A series samples (a, b) and B series samples (c, d); (b, d) are enlarged views of the local region.

4.4. Complete diamagnetism (Meisner effect)

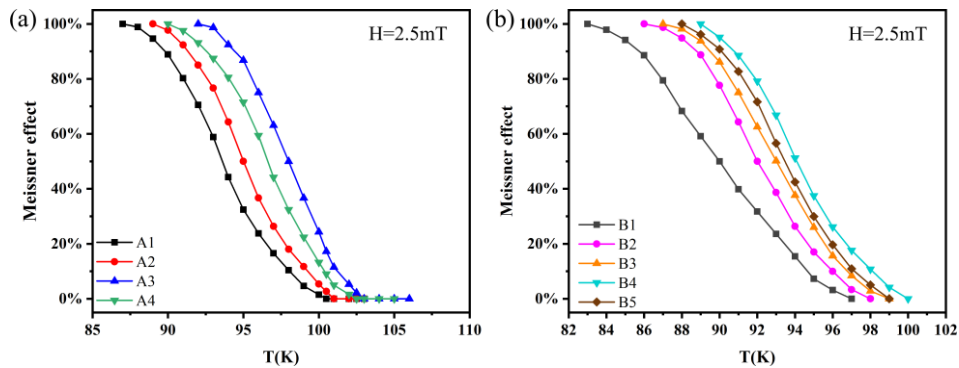

**Figure 7.** (a) DC magnetization data of A series samples; (b) DC magnetization data of B series samples.

Figure 7 shows the DC magnetization data for the A and B series samples. The horizontal axis represents temperature, and the vertical axis represents the percentage change in material's complete diamagnetic response, expressed as the Meissner effect. From the DC magnetization data, it can be observed that all samples exhibit complete diamagnetism (Meissner effect). As the temperature increases, the complete diamagnetic response (Meissner effect) weakens and eventually disappears, consistent with the literature [72, 73]. From Figure 7(a), it can be seen that the complete diamagnetic response (Meissner effect) of the pure B(P)SCCO sample A1 completely disappears at 100.5K, while for the inhomogeneous doped samples A2, A3, and A4 with doping contents of 0.1wt.%, 0.15wt.%, and 0.2wt.%, the complete diamagnetic response (Meissner effect) disappears at temperatures of 101K, 103K, and 102.5K, respectively. Figure 7(b) shows that at a temperature of 97K, the complete diamagnetic response (Meissner effect) of the pure B(P)SCCO sample B1 completely disappears, while for the inhomogeneous phase-doped samples B2, B3, B4, and B5 with doping contents of 0.1wt.%, 0.15wt.%, 0.2wt.%, and 0.25wt.%, the complete diamagnetic response (Meissner effect) disappears at temperatures of 98K, 99K, 100K, and 99K, respectively. In both the A and B series samples, the inhomogeneous phase-doped samples exhibit an increased complete diamagnetic response (Meissner effect) compared to the pure B(P)SCCO samples, indicating that inhomogeneous phase doping effectively enhances the complete diamagnetic response (Meissner effect) of the B(P)SCCO samples.

**4. Discussion**

we introduced green light emitting GaN p-n junction particles as inhomogeneous phase inclusions into B(P)SCCO superconductors, forming smart meta-superconductors (SMSCs). We investigated the effect of the doping content of the p-n junction luminescent particles on the superconducting properties of B(P)SCCO superconductors with different matrix sizes. The following points need further elaboration:

5.1. The enhancement effect of green light p-n junction luminescent particle doping.

Compared to pure B(P)SCCO samples, the doping of GaN p-n junction particles can indeed enhance the critical transition temperature ($T_C$), critical current ($J_C$), and perfect diamagnetism (Meissner effect) of B(P)SCCO superconductors, and the enhancement effect varies with the content of non-uniform phase doping in a dome-shaped manner. In the A series samples, the doping of inhomogeneous phase increased the zero-resistance transition temperature $T_{C,0}$ of the samples by 2K and increased the critical current $J_C$ by 25%; in the B series samples, the doping of inhomogeneous phase increased the zero-resistance transition temperature $T_{C,on}$ by 3K, increased the starting transition temperature $T_{C,0}$ by 2K, and increased the critical current $J_C$ by 35%.

5.2. Adjusting the matrix size of B(P)SCCO superconductors can regulate the enhancement ef-fect of superconducting properties.

Comparing the A and B series samples, the smaller the particle size of the raw material, the smaller the size of the generated B(P)SCCO polycrystalline particles, and the more grain boundaries, resulting in lower critical temperature ($T_C$) and critical current ($J_C$) for the B series samples compared to the A series samples. In addition, for the A series samples prepared from raw materials with a particle size of 30μm, the optimal content of inhomogeneous phase doping is 0.15 wt.%, while for the B series samples prepared from raw materials with a particle size of 5μm, the optimal content is 0.2 wt.%, further enhancing the superconducting properties of B(P)SCCO superconductors. This indicates that in our designed B(P)SCCO smart meat-superconductor structure, the smaller the particle size of the B(P)SCCO matrix, the more inhomogeneous phase content it can accommodate, resulting in better performance enhancement.

5.3. Comparison of chemical doping and luminescent inhomogeneous phase doping.

Chemical doping is a commonly used method to enhance superconducting properties. In previous studies, some researchers attempted to enhance the superconducting properties of BSCCO superconductors through element substitution. The most successful method so far is the substitution of $Pb^{2+}$ for $Bi^{3+}$ to enhance the stability of the Bi-2223 high-temperature phase and increase its content, but it cannot increase the critical temperature ($T_C$). Other elements such as Al, Cs, and Ce can only effectively increase the critical current density ($J_C$). Another group of researchers used compound doping, such as $Al_2O_3$, $SnO_2$, $ZrO_2$, MgO, etc. Studies have shown that compound doping can also only increase the critical current density ($J_C$). In this study, we introduced GaN p-n junction luminescent particles as inhomogeneous phase inclusions into B(P)SCCO superconductors. Through experiments, we found that inhomogeneous phase doping simultaneously enhanced the critical temperature ($T_C$), critical current density ($J_C$), and complete diamagnetism of B(P)SCCO superconductors. Compared to chemical doping, which only increases the critical current density ($J_C$), this method has significant advantages. This method of enhancing superconductivity through the electroluminescence effect of luminescent particles provides a new approach to explore superconductor modification.

5.4. Universality of enhancing superconducting performance through luminescent inhomogeneous phase doping.

In previous experiments, our team introduced $Y_2O_3{:}Eu^{3+}$+Ag electroluminescent particles as inhomogeneous phase dopants into B(P)SCCO high-temperature superconductors to enhance their superconducting properties. The optimal doping content of $Y_2O_3{:}Eu^{3+}$+Ag electroluminescent particles was found to be 0.2 wt.%, which increased the zero-resistance transition temperature ($T_{C,0}$) by 1K and the onset transition temperature ($T_{C,\,on}$) by 1.1K compared to the same batch of pure B(P)SCCO superconductors [64,65]. Now, with green GaN p-n junction luminescent particles, the enhancement effect on the onset transition temperature ($\Delta T_{C,\,on}$) of the same batch of samples reached 2K. It can be observed that the effect of these electroluminescent inhomogeneous phases is not significantly influenced by the material composition but depends on their luminescent performance under an electric field. The use of electroluminescent inhomogeneous phase doping to improve superconducting performance is a universally existing phenomenon.

5.5. Enhancement effect of p-n junction luminescent particle doping on different supercon-ductors.

We also introduced red AlGaInP p-n junction particles as inhomogeneous phase dopants into MgB2 conventional superconductors, increasing the critical temperature to 39K and $\Delta T_{C,\,on}$ to 0.8K [67]. Now, with green GaN p-n junction luminescent particle doping in B(P)SCCO high-temperature superconductors, the critical transition temperature is also increased, and $\Delta T_{C,\,on}$ reaches 2K. Whether for conventional or high-temperature superconductors, the use of p-n junction electroluminescent particle doping to enhance superconductivity is feasible and universal.

5.6. The physical origin of the increase in critical transition temperature

Through experiments, we observed that both $Y_2O_3{:}Eu^{3+}$+Ag rare earth oxide luminescent particles and p-n junction luminescent particles, as inhomogeneous phase dopants in superconductors, can enhance the superconductivity to varying degrees. However, there is currently no theoretical model that can well explain this phenomenon in accordance with experimental results. Therefore, based on years of experimental research, we propose the idea that the coupling between the evanescent waves generated by electroluminescence and the superconducting electrons enhances the critical transition temperature ($T_C$) of the superconductor. By introducing p-n junction luminescent inhomogeneous phases, the photons generated by electroluminescent particles interact with some superconducting electrons, resulting in the formation of surface plasmas. These evanescent waves promote the smooth transmission of superconducting electrons with the same energy, thereby facilitating electron interactions in the surface plasma system. The injection of energy improves the formation of electron pairs, enhances the superconducting behavior of the material, increases the critical transition temperature, and leads to the formation of the superconductivity enhancement effect. At the same time, inhomogeneous phase doping can also cause impurity effects, which compete with electroluminescence effects. With the increase in inhomogeneous phase content, the impurity effect is enhanced, and the enhancement of superconductivity shows a dome-shaped variation. Therefore, under the co-evolution of electroluminescence effects and impurity effects, the smart meta-superconductivity of electroluminescent inhomogeneous phase-doped superconductors is formed.

## 5. Conclusions

In this paper, we constructed B(P)SCCO smart meta-superconductors (SMSCs) by introducing green light GaN p-n junction particles as inhomogeneous phases into B(P)SCCO superconducting particles. Two different series of samples were prepared using solid-state sintering method with two different particle size raw materials. The critical temperature ($T_C$), critical current ($J_C$), and complete diamagnetism (Meissner effect) of the samples were measured using four-probe method, I-V curve measurement, and DC magnetization measurement. The influence of introducing green light GaN p-n junction inhomogeneous phases on the superconducting properties of B(P)SCCO samples was investigated. The experimental results showed that by optimizing the doping of luminescent inhomogeneous phases, the critical transition temperature ($T_C$) can be increased by 3K, the critical current density ($J_C$) can be increased by 35%, and the complete diamagnetism (Meissner effect) can be significantly enhanced. Through systematic studies, we found that the combination of electroluminescence effect and impurity effect caused by traditional inhomogeneous doping phase can change the superconducting properties of superconductors. The stronger the electroluminescence effect of luminescent inhomogeneous phases, the better the enhancement of superconducting properties. This study provides valuable insights and guidance for the optimization design and application of superconducting materials. Future research can further explore suitable doping materials with excellent performance and investigate the physical mechanisms of enhancing superconducting properties, thus further promoting the exploration and application of superconducting materials.

**Author Contributions:** Conceptualization, methodology, X.Z.; software, Q.H., H.C. and D.C.; validation, Q.H., H.C. and C.S.; formal analysis, Q.H., H.C., C.S., D.C., Y.Q. and M.S.; investigation, Q.H., H.C., C.S., D.C., Y.Q. and M.S.; resources, X.Z.; data curation, Q.H., H.C. and C.S.; writing—original draft preparation, Q.H.; writing—review and editing, Q.H. and X.Z.; visualization, Q.H., Q.Y. and M.S.; supervision, X.Z.; project administration, X.Z.; funding acquisition, X.Z. All authors have read and agreed to the published version of the manuscript.

**Funding:** This research was supported by the National Natural Science Foundation of China for

Distinguished Young Scholar under Grant No. 50025207.

**Data Availability Statement:** The data presented in this study are available on reasonable request from the corresponding author.

**Conflicts of Interest:** The authors declare no conflict of interest.